\documentclass[a4paper]{article}
\usepackage{hyperref}
\usepackage{makeidx}
\usepackage{color}
\makeindex
\usepackage{graphicx} 
\usepackage{bm}
\usepackage{amssymb}
\usepackage{amsmath}
\usepackage{bm}
\usepackage{mathrsfs}

\def\rhpDE{{\rho_{\hspace{-.15em}D\hspace{-.15em}E}}}

\def\start{{\star\star}}
\def\bcdot{{\circ}}
\def\bcdots{{\circ\circ}}
\def\bcdott{{\circ\circ\circ\circ}}

\def\bullett{{\bullet\bullet}}

\def\aaa{\mathfrak{a}}
\def\AAA{\mathfrak{A}}

\def\bbb{\mathfrak{b}}
\def\BBB{\mathfrak{B}}

\def\ooo{\mathfrak{o}}

\def\C{\mathbb{C}}
\def\Z{\mathbb{Z}}

\def\TTT{{\mathfrak T}}
\def\ttt{{\mathfrak t}}

\def\td{d_\www}

\def\vvv{\mathfrak{v}}
\def\VVV{\mathfrak{V}}

\def\sss{\mathfrak{s}}

\def\SSS{{\mathfrak S}}

\def\LL{{\cal L}}

\def\LLL{{\mathfrak L}}

\def\HH{{\cal H}}

\def\HHH{{\mathfrak H}}

\def\aaa{{\mathfrak a}}
\def\AAA{{\mathfrak A}}

\def\BBB{{\mathfrak B}}
\def\bbb{{\mathfrak b}}

\def\C{\mathbb{C}}
\def\CCC{{\mathfrak C}}
\def\ccc{{\mathfrak c}}

\def\Varepsilon{{\mathcal E}}

\def\eee{{\mathfrak e}}

\def\I{\mathscr{I}}

\def\M{{\cal M}}
\def\MM{\mathscr{M}}
\def\MMM{\mathfrak{M}}

\def\R{\mathbb{R}}

\def\RRR{\mathfrak{R}}

\def\vomega{{\mathfrak w}}
\def\www{{\mathfrak w}}
\def\WWW{{\mathfrak W}}

\def\kh{\kappa\hbar}
\def\Gh{G\hbar}

\def\LLvac{\LL_{\hspace{-.1em}\textrm{vac}}}

\def\TM{{T\hspace{-.2em}\M}}
\def\TsM{{T^*\hspace{-.2em}\M}}
\def\TMM{{T\hspace{-.2em}\MM}}
\def\TsMM{{T^*\hspace{-.2em}\MM}}

\def\GSO{{G_{\hspace{-.1em}S\hspace{-.1em}O}}}
\def\SO{{\hspace{-.1em}S\hspace{-.1em}O\hspace{-.1em}}}

\def\stars{{\star\star}}

\begin{document}
\title{Quantum effect on black holes and cosmological constant problem}
\author{Yoshimasa Kurihara\footnote{yoshimasa.kurihara@kek.jp}
\\
{\it\small High Energy Accelerator Organization (KEK), 
Tsukuba, Ibaraki 305-0801, Japan}
}
\date{}
\maketitle
\begin{abstract} 
We propose a quantum gravity equation owing to the geometrical quantization of general relativity, namely the Schr\"{o}dinger--Einstein equation.
Quantum effects of a Schwarzschild black hole are demonstrated by solving the quantum equation requiring a stationary phase; the consistent result is obtained using the Einstein--Brillouin--Keller (EBK) quantization condition. 
We solve the Schr\"{o}dinger--Einstein equation around the classical solution of the McVittie--Thakurta metric. 
This solution simultaneously describes a system with Schwarzschild's black holes and a scalar field. 
Moreover, we investigate a possible interplay between quantum black holes and a scalar field.
The number density of black holes in the universe is obtained by applying statistical mechanics to a system consisting of black holes and a scalar field.
A possible solution to the cosmological constant problem is proposed from a statistical perspective.
\end{abstract}
\maketitle
\section{Introduction}\label{intro}
From the cosmic scale to the sub-atomic scale, our understanding of nature has increased considerably in recent years.
Phenomena on the scale of the Universe are described using the theory of general relativity, whereas the quantum theory of gauge fields governs the microscopic scale.
However, the unification of these two fundamental theories remains to be achieved.
The development of a quantum theory of gravity in four-dimensional space-time is one of the most significant challenges of modern fundamental physics.

The old quantum theory played an essential role in developing our understanding of atomic systems and in the growth of modern quantum mechanics.
For example, the Bohr--Sommerfeld quantum condition successfully predicted the hydrogen atom's allowed (discrete) energy states.
This success led to the concept of a particle (electron) field, as described by de\hspace{.2em}Broglie\cite{10.1051/anphys/192510030022}.
Whereas the Bohr--Sommerfeld quantum condition applies only to systems undergoing periodic motion, its applicability was extended to systems in non-periodic motion by Einstein\cite{Einstein:422242} and Brillouin\cite{brillouin1926remarques}.
Later, Keller\cite{KELLER1958180} further developed the theory by adding a correction term to the quantization condition to account for the boundary conditions.
This effect is now known as Einstein--Brillouin--Keller (EBK) quantization.
Keller's correction term was refined by Maslov\cite{maslov75theory,maslov1965wkb} and Arnol'd\cite{Arnol'd1967} in the context of symplectic geometry by introducing the so-called Keller--Maslov--Arnol'd (KMA) correction term.

The symplectic structure of general relativity and EBK quantization for the vacuum solution of the Einstein equation were previously discussed in detail by the current author\cite{Kurihara_2020}, and a prequantization bundle and a prequantum Hilbert space of general relativity were proposed.
The geometric-quantization method involves a quantum equation of gravity (the Schr\"{o}dinger--Einstein equation) expressed in terms of quantum operators that satisfy canonical commutation relations.
The zeroth-order solution of the Schr\"{o}dinger--Einstein equation amounts to a classical vacuum solution that, in turn, yields approximate quantum wave functions beyond the vacuum solutions.
The present study discusses quantum effects in Schwarzschild black holes with a scalar field.

This report is organized as follows:
In \textbf{section \ref{CGR}}, classical general relativity is briefly reviewed using the terminology employed in previous studies\cite{Kurihara:2016nap,Kurihara_2020,doi:10.1063/1.4990708,doi:10.1140/epjp/s13360-021-01463-3}.
Two formalisms of quantum general relativity, EBK quantization and geometric quantization, are then discussed in \textbf{section \ref{QM}}.
\textbf{Section \ref{ASBH}} concerns applications of quantum general relativity to a Schwarzschild black hole and to a system consisting of a black hole and a scalar field.
\textbf{Section\ref{CCP}} builds upon \textbf{section \ref{ASBH}}, treating the Universe as being in thermal equilibrium of black holes and a scalar field.
A possible solution to the so-called ``cosmological constant problem'' is also discussed in \textbf{section \ref{CCP}}.
Finally, the results are summarized in \textbf{section  \ref{SUMMARY}}.

\section{Classical theory of general relativity}\label{CGR}
\subsection{Notations}
A four-dimensional pseudo-Riemannian manifold $(\MM,\bm{g})$ with $GL(4,\R)$ symmetry is introduced as a model of the universe.
Its tangent and cotangent bundles are denoted $\TMM$ and $\TsMM$.
This report uses  Fraktur letters ($\AAA,\aaa,\BBB,\bbb,\CCC,\ccc\cdots$) for differential forms in $\TsMM$.
Spaces of tangent- and cotangent vectors (forms) are denoted as $V^p(\TMM)$ and $\Omega^p(\TsMM)$, where $p$ is the rank of tensors.

A line element is defined using the metric tensor as $ds^2=g_{\mu_1\mu_2}(x)\hspace{.1em}dx^{\mu_1}\otimes dx^{\mu_2}$.
(The Einstein convention for repeated indices is used throughout this study.)
It is always possible to find a vanishing affine connection frame at any point $p\in\MM$, the Einstein equivalence principle.
A local manifold with a vanishing affine connection has Poincar\'{e} symmetry $I\hspace{-.2em}S\hspace{-.1em}O(1,3)=SO(1,3)\ltimes T{(4)}$.
This local manifold is referred to as the local Lorentz manifold and is denoted $\M_p$. 
The trivial bundle $\M:=\bigcup_p\M_p$ is introduced  in base manifold $\MM$ with the structure group $I\hspace{-.2em}S\hspace{-.1em}O(1,3)$.
The tangent bundle $\TM:=\bigcup_p\TM_p$ and cotangent bundle $\TsM:=\bigcup_p\TsM_p$ are also introduced.
We adopt the metric tensor $[{\bm \eta}]=\textrm{diag}(1,-1,-1,-1)$ in $\M$.
Standard local-coordinate vectors in $\TM$ is represented as $\partial_a:=\partial/\partial x^a$.
In our representation, the base manifold of vectors and forms $\MM$ or $\M$ are distinguished by Greek suffixes in $\MM$ or Roman suffixes in $\M$. 
Map $\Varepsilon:{\bm g}\mapsto{\bm \eta}$ is represented using the standard basis on any chart of $\MM$ as
\begin{align*}
\eta^{ab}&=\Varepsilon^a_{\mu_1}(x)\Varepsilon^b_{\mu_2}(x)g^{\mu_1\mu_2}(x),
\end{align*}
where $\eta^{ab}=[{\bm \eta}]^{ab}$ and $g^{\mu\nu}=[{\bm g}]^{\mu\nu}$.
Function $\Varepsilon^a_{\mu}(x)=[{\bm \Varepsilon}(x)]^a_{\mu}$ is referred to as a vierbein, that induces isomorphism $\TMM\cong \TM$.
Orthogonal basis is obtained from those in $\TMM$ and $\TsMM$, respectively, as $\partial_a:=\Varepsilon_a^\mu\partial_\mu$ and $\eee^a:=\Varepsilon^a_\mu dx^\mu$, where $\Varepsilon_a^\mu:=[{\bm \Varepsilon}^{-1}]_a^\mu$.
They are dual to each other such that $\eee^a\hspace{.1em}\partial_b=\Varepsilon^a_{\mu_1}\Varepsilon_b^{\mu_2}dx^{\mu_1}\hspace{.1em}\partial_{\mu_2}=\Varepsilon^a_{\mu}\hspace{.1em}\Varepsilon_b^{\mu}=\delta^a_b$.
Standard basis $\eee^a$, namely the virbein form, is a vector in tangent- and cotangent space as $\eee\in\Omega^1(\TsM)\otimes{V}^1(\TM)$.
The spin connection form $\www\in\Omega^1(\TsM)\otimes{V}^2(\TM)\otimes\sss\ooo(1,3)$ is a Lie algebra valued one-form object and  represented in the standard basis as $\www^{ab}=\omega^{~a}_{\mu~c}\hspace{.1em}\eta^{cb}dx^\mu$.
We note that it is antisymmetric concerning tensor indices as $\www^{ab}=-\www^{ba}$.

The torsion two-form $\TTT^a$ is defined owing to the covariant differential such that $\TTT^a:=\td\eee^a=d\eee^a+\www^a_{~b}\wedge\eee^b$, where $\td$ is a covariant differential with respect to the structure group of $SO(1,3)$.
This report adopts the convention whereby Roman dummy indices are represented with a circle or a star when the Einstein convention pairing is trivial.
However, Greek dummy indices are not abbreviated in this manner.
The curvature two-form is defined as $\RRR^{ab}:=d\www^{ab}+\www^a_{~\bcdot}\wedge\www^{\bcdot b}$, which is a Lorentz-tensor-valued two-form object $\RRR^{ab}\in V^2(\TM)\otimes\Omega^2(\TsM)\otimes\sss\ooo(1,3)$.
Multiple circles in an expression signify that pairing must be done in the left-to-right direction for both upper and lower indices.
A $GL(4,\R)$ invariant volume form is provided as
\begin{align*}
\vvv&=\frac{1}{4!}\epsilon_{\bcdots\bcdots}\Varepsilon^\bcdot_{\mu_1}\Varepsilon^\bcdot_{\mu_2}
\Varepsilon^\bcdot_{\mu_3}\Varepsilon^\bcdot_{\mu_4}
dx^{\mu_1}\wedge dx^{\mu_2}\wedge dx^{\mu_3}\wedge dx^{\mu_4},\\
&=\mathrm{det}[{\bm\eta}]\hspace{.1em}dx^{0}\wedge dx^{1}\wedge dx^{2}\wedge dx^{3}.
\end{align*}
where $\epsilon_{abcd}$ is a completely antisymmetric tensor  in $\TM$ with $\epsilon_{0123}=1$.
A two-dimensional surface form is defined as $\SSS_{ab}:=\epsilon_{ab\bcdots}\eee^\bcdot\wedge\eee^\bcdot/2$, which is a two-dimensional surface perpendicular to both $\eee^a$ and $\eee^b$.
We note that the volume form can be written using the surface form as $\vvv=-\epsilon^{\bcdots\bcdots}\SSS_\bcdots\wedge\SSS_\bcdots$/4!\hspace{.1em}.

\subsection{Lagrangian and Hamiltonian formalisms}
The Einstein--Hilbert gravitational Lagrangian $\LLL_G$ is defined as
\begin{align}
\LLL_G
&:=\frac{1}{\hbar\kappa}\left(
\RRR^\bcdots\wedge\SSS_\bcdots
-\Lambda_{\textrm{c}}\vvv
\right),\label{Lag}
\end{align}
where $\kappa=4\pi G\approx1.03829\hspace{.1em}\textrm{s}^2\textrm{m}^{-1}\textrm{kg}^{-1}$ is the Einstein gravitational constant and  $G$ is the Newtonian gravitational constant, and $\Lambda_{\textrm{c}}$ is the cosmological constant.
Whereas light velocity is set to unity ($c=1$), the other physical constants $\kappa$ and $\hbar$ are written out explicitly in this study.
By this convention, we express fundamental parameters such as the Planck length $l_p=\sqrt{\kappa\hbar}$, the Planck time $t_p=\sqrt{\kappa\hbar}$ and the Planck mass $m_p=\sqrt{\hbar/\kappa}$.
We note that the gravitational Lagrangian $\LLL_G$ is dimensionless by this convention. 
The Einstein equation and torsion-less condition are provided as an Euler--Lagrange equation of motion concerning the vierbein- and spin forms such that:
\begin{align}
\epsilon_{a\bcdots\bcdot}\RRR^\bcdots\wedge\eee^\bcdot=0~~&\textrm{and}~~\TTT^a=0.\label{Eeq}
\end{align}
The cosmological constant is henceforth assigned a zero value.
\textbf{Section \ref{CCP}} will discuss the possible origin of the cosmological constant.

The Hamiltonian formalism is constructed by identifying the fundamental forms as $(\www,\SSS)$\cite{Kurihara:2016nap,Kurihara_2020,doi:10.1140/epjp/s13360-021-01463-3}.
When spin form $\vomega$ is identified as the general \textit{configuration} variable, canonical \textit{momentum} $\MMM$ is provided as
\begin{align*}
\MMM_{ab}:=\frac{\delta{\LLL}_G}{\delta\left(d\vomega^{ab}\right)}=\frac{1}{\kappa\hbar}\SSS_{ab}.
\end{align*}
This result is consistent with the choice of phase space $(\www,\SSS)$ owing to the principal co-Poincar\'{e} bundle\cite{doi:10.1063/1.4990708}.
The classical Hamiltonian form can be obtained from the Lagrangian form using the Legendre transformation as
\begin{align}
\HHH_G&:=
\frac{1}{2}\MMM_\bcdots\wedge d\www^\bcdots-\frac{1}{\hbar}{\LLL_G}=
-\frac{1}{2\kappa\hbar}
\vomega^{\star}_{~\bcdot}\wedge\vomega^{\bcdot\star}\wedge\SSS_{\star\star}\hspace{.2em}.\label{CHG}
\end{align}
We note that the Hamiltonian form has no physical dimensions in this study and does not have a local ${SO}(1,3)$ invariance.
The Einstein equation and torsion-less condition can be obtained as canonical equations of motion, as expected.

The Poisson bracket is introduced in the Lorentz-covariant formalism as
\begin{align*}
\left\{\aaa,\bbb\right\}^{~}_\textrm{PB}&:=
\frac{\delta\aaa}{\delta\vomega^\bcdots}\wedge\frac{\delta\bbb}{\delta\SSS_\bcdots}-
\frac{\delta\bbb}{\delta\vomega^\bcdots}\wedge\frac{\delta\aaa}{\delta\SSS_\bcdots},
\end{align*}
where $\aaa\in\Omega^p(\TsM)$ and $\bbb\in\Omega^q(\TsM)$ for $0\leq p,q\in\Z$.
The Poisson brackets for the fundamental forms are provided as
\begin{align}
\begin{array}{cl}
\left\{\vomega^{a_1a_2},\vomega^{a_3a_4}\right\}_\textrm{PB}&=
\left\{\SSS_{b_1b_2},\SSS_{b_3b_4}\right\}_\textrm{PB}~=~0,\\
\left\{\vomega^{a_1a_2},\SSS_{b_1b_2}\right\}_\textrm{PB}&=
\delta^{[a_1}_{b_1}\delta^{a_2]}_{b_2},
\end{array}
\label{pb}
\end{align}
where $\delta^{[a_1}_{b_1}\delta^{a_2]}_{b_2}=\delta^{a_1}_{b_1}\delta^{a_2}_{b_2}-\delta^{a_2}_{b_1}\delta^{a_1}_{b_2}$.
The Hamiltonian form is a generator of a total derivative for a given form  object; the Poisson bracket between the fundamental- and Hamiltonian forms yields owing to the canonical equations of motion such that
\begin{align*}
\epsilon_{a\bcdots\bcdot}\left\{\vomega^\bcdots,\HHH_G\right\}_\textrm{PB}\wedge\eee^\bcdot&=
-\epsilon_{a\bcdots\bcdot}\vomega^\bcdot_{~\star}\wedge\vomega^{\star\bcdot}\wedge\eee^\bcdot
=\epsilon_{a\bcdots\bcdot}d\vomega^\bcdots\wedge\eee^\bcdot,
\end{align*}
and
\begin{align*}
\left\{\SSS_{ab},\HHH_G\right\}_\textrm{PB}&=
-\left(-\eta_{b\bcdot}\vomega^\bcdots\wedge\SSS_{\bcdot a}\right)=d\SSS_{ab}.
\end{align*}

\section{Quantization methods}\label{QM}
The current author discusses\cite{Kurihara_2020} the EBK quantization condition for the vacuum solutions of the Einstein equation in terms of geometrical quantization based on the symplectic structure of the space-time manifold.
This section summarizes the quantization condition of the vacuum solution. 
It presents the quantum equation of motion owing to geometrical quantization under the context of general relativity.
We also introduce the quantum Hilbert space in the prequantization submanifold and define the Schr\"{o}dinger--Einstein equation.
%
%
\subsection{EBK quantization}
This section considers a system of $N$ classical particles with Hamiltonian $H(q_i,p_i)$, $(i=1,\cdots,N)$.
The Hamiltonian is assumed to be completely integrable.
The Liouville form $\sss_i\hspace{-.2em}=\hspace{-.2em}p_idq_i$ induces the symplectic manifold $(\R^{2N},d\sss_i)$.
The submanifold $\LL\subset(\R^{2N},d\sss_i)$ is hence referred to as the lagrangian submanifold when it satisfies $\WWW|_\LL=0$ and $\textrm{dim}(\LL)=N$.
If separation of the phase-space variables is impossible, the particles do not form a closed orbit.
Einstein, Brillouin, and Keller extended the Bohr--Sommerfeld quantization condition to a non-closed orbit case such that:
\begin{align*}
S^\textrm{EBK}_k=\frac{1}{2\pi\hbar}\oint_{\Gamma_k}\sum_{i=1}^Np_idq_i=
n_k+\frac{\mu_k}{4}, ~~&\textrm{for}~~k=1,\cdots,N,
\end{align*}
where $0\leq\mu_i\in\Z$ is the index of the $k$th variable obtained owing to the boundary conditions.
Contour integrals are performed along the homotopy-independent closed-circles of each  particle in the lagrangian submanifold. 
A contour is denoted $\Gamma_k$ and is not necessarily a classical closed orbit.
The constant $\mu_k$ is referred to as the KMA index in this study. 

Classical general relativity with the principal co-Poincar\'{e} bundle induces symplectic manifold $(\sigma\otimes\varpi, d\SSS_\bcdots\wedge d\www^\bcdots)$, where $\varpi$ and $\sigma$ are, respectively, spaces of the spin connection- and surface forms as vacuum solutions of the Einstein equation.  
The EBK quantization condition is formulated for the prequantization submanifold as\cite{Kurihara_2020}
\begin{align}
\int_{\LLvac}\SSS_\bcdots\wedge d\www^\bcdots&=\int_{\LLvac}\left(\LLL_G+\HHH_G\right)=n.
\label{EBKGRvac}
\end{align}
where $\LLvac$ is the lagrangian submanifold that gives $\LLL_G|_{\LLvac}\hspace{-.2em}=\hspace{-.2em}0$, and $0\hspace{-.2em}\leq\hspace{-.2em}n\in\Z$ is a quantum number.
After integration, the vacuum energy ${E}_{{\hspace{-.1em}\textrm{vac}}}$ is given by
\begin{align}
\int_{\Sigma}\HHH_G&={E}_{{\hspace{-.1em}\textrm{vac}}}=
n+\frac{[H^4(\M_5,\Z)]}{\nu},\label{Evac}
\end{align}
where $\Sigma\subset\M$ is an appropriate submanifold of space-time manifold $\M$, and $\nu\in\R$ is a constant to be determined from a solution of the quantum equation.
Here, the four-dimensional space-time manifold is embedded in a five-dimensional manifold $\M_5$. 
We note that the Lagrangian form in the integral (\ref{EBKGRvac}) is, simultaneously, the fundamental-form valued four-form in $\sigma\otimes\varpi$ and also the standard four-form $\LLL_G\in\Omega^4(\TsM)$.
These two Lagrangian forms are equated under the homomorphism $\LLvac\simeq\TsM$ discussed in Ref.\cite{Kurihara_2020}.
The KMA index arises from integrating the Lagrangian form as the second Chern class\cite{doi:10.1063/1.4990708}. 

%
%
\subsection{Geometric quantization}
The vacuum energy spectrum has been considered in terms of EBK quantization without a quantum state vector.
This section introduces the state vector and discusses its physical interpretation.

In our method, a target of quantization is not the space-time manifold itself; thus, the space-time coordinate $x^\mu$ is not an operator\cite{nakanishi1990covariant,NakanishiSK2009}.
Instead, the classical vierbein $\Varepsilon^{(c)}$ is quantized (or, equivalently, ${\bm g}^{(c)}$), which is provided as a solution of the Einstein equation.
In classical general relativity, the geometric Riemannian-metric tensor ${\bm g}^{(g)}$ is equated with the solution of the classical Einstein equation, such as ${\bm g}^{(g)}={\bm g}^{(c)}$, i.e., the Einstein equivalence principle.
At a quantum level, this relation is not simply fulfilled.
The geometric metric tensor is given as the \textit{expected value} (as defined below) of the quantum metric tensor ${\bm g}^{(g)}=\langle\Psi| {\bm g}^{(q)}|\Psi\rangle$, where $|\Psi\rangle$ is a state vector of quantum general relativity.
A space with metric tensor provided as the expected value of a stochastic process was discussed in \cite{Kurihara:2016gyt} and the unitarity of the transition matrix was shown in \cite{doi:10.1140/epjp/s13360-021-01463-3}.

Under the context of geometric quantization, the Poisson bracket (\ref{pb}) is replaced by the commutation relation of operators as
\begin{align}
\begin{array}{cl}
\left[\widehat{\vomega}^{a_1a_2}(x),\widehat{\vomega}^{b_1b_2}(y)\right]&=
\left[\widehat{\SSS}_{a_1a_2}(x),\widehat{\SSS}_{b_1b_2}(y)\right]=0,\\
\left[\widehat{\vomega}^{a_1a_2}(x),\widehat{\SSS}_{b_1b_2}(y)\right]&=i\kh\hspace{.1em}\delta^{(4)}(x-y)
\delta^{[a_1}_{b_1}\delta^{a_2]}_{b_2}.
\end{array}
\label{QCCR}
\end{align}
We use the representation $\hat\bullet$ for the operator corresponding to physical quantity $\bullet$ in this study.
The corresponding prequantum operators are obtained (see, e.g., Ref.\cite{nair2005quantum}) as
\begin{align}
\widehat{\www}^{ab}={\hspace{.9em}i\kh}\left[\frac{\delta~~}{\delta\SSS}\right]^{ab}+\www^{ab}
~~&\textrm{and}~~
\widehat{\SSS}_{ab}={-i\kh}\left[\frac{\delta~~}{\delta\www}\right]_{ab},
\label{CCRSW1}
\end{align}
which are the reducible representation.
Direct calculations confirm that the operators (\ref{CCRSW1}) satisfy commutation relations (\ref{QCCR}).
When real-polarization is exploited as in Ref.\cite{Kurihara_2020}, the state vector is give as a function of vacuum solutions for the spin connection; thus, the irreducible representation of the spin-form operator is provided as $\widehat{\www}^{ab}=\www^{ab}$.

Follwoing Ref.\cite{doi:10.1140/epjp/s13360-021-01463-3}, we discuss the functional space of the spin connection here.
Classically, a spin connection is obtained as a solution of the Einstein equation, which is a nonlinear first-order differential equation.
This section treats a homogeneous equation for a pure gravitational equation without any gauge or matter fields.
When classical solution $\www^{(c)}$ is obtained, the Lorentz-transformed form ${\GSO}(\www^{(c)})$ is also the solution of the equation; thus, quotient space $\varpi:=\widetilde\varpi/{\GSO}(\widetilde\varpi)$ is introduced, where $\widetilde\varpi$ is a set of solutions of the classical Einstein equation. 
When local Lorentz manifold $\M$ has a Euclidean metric, $\varpi$ has the Sobolev norm $L^p_k(\www^{(c)})$ such that:
\begin{align*}
\left\|\www^{(c)}\right\|_{L^p_k}:=\left(
\sum_{0\leq i_1+\cdots+i_n\leq k}\int_{\Sigma_\M}
\left| 
{\partial_{i_1}\cdots\partial_{i_n}} \omega^{(c)}
\right|^p\vvv
\right)^{1/p},
\end{align*}
where $\Sigma_\M$ is a compact subset of $\M$.
The summation is performed over all possible combinations $(i_1,\cdots,i_n)$ with $0\leq i_1+\cdots+i_n\leq k$, where $0\leq i_j\in\Z$.  
In our case, dimension  $n$ of manifold $\M$ is $n=4$.
The completion of $\varpi$ with respect to the Sobolev norm makes the Sobolev space, denoted as $L^p_k(\varpi)$.
The Einstein equation, however, is not elliptic even in Euclidean space, because the classical equation is invariant under the infinite groups $GL(4,\R)$ and $\SO(4)$.
However, the quantum Lagrangian, which includes (local) gauge-fixing terms, can be expected to be locally elliptic; thus, functional analysis can be applied locally around a gauge-fixing point.
In this case, $L^2_k(\varpi)$ is a Hilbert space with respect to an inner product defined as
\begin{align*}
\left(\www^{(c)},{\www^{(c)}}'\right)_{L^2_k(\varpi)}:=
\sum_{0\leq i_1+\cdots+i_n\leq k}\int_{\Sigma}
\left(\partial_{i_1}\cdots\partial_{i_n}\omega^{(c)}
\right)\cdot
\left(\partial_{i_1}\cdots\partial_{i_n}{\omega^{(c)}}'
\right)\vvv.
\end{align*}
A dot-product is defined using a metric tensor in functional space.
Hilbert space $L^2_0(\varpi)$ is denoted as $\HH^{(c)}$.

Functional space $\varpi$ is a subset of $\HH^{(c)}$ as $\varpi\subset\HH^{(c)}\subset L^2$.
The Hodge-dual space of $\varpi$, denoted as $\tilde{\varpi}$, is simply referred to as the dual space of $\varpi$.
When $\omega\in\varpi$ is square-integrable, dual spin-connection $\tilde{\omega}\in\tilde{\varpi}$ is also square-integrable; thus, the Gel'fand triple becomes  $\varpi\subseteq\HH^{(c)}\subseteq\tilde{\varpi}\subset L^2$.
The coefficient function of the spin-  and dual spin-form are denoted as 
\begin{align*}
\omega^{\hspace{.3em}ab}_{\mu}\in\varpi ~~&\textrm{and}~~ 
\tilde\omega^{\hspace{.3em}ab}_{\mu}:=\frac{1}{2}\eta^{a\star}\eta^{b\star}\hspace{.2em}
\epsilon_{\stars\bcdots}\hspace{.2em}\omega^{~\bcdots}_{\mu}\in\tilde\varpi
\end{align*}
A standard bilinear form is defined as
\begin{align*}
\langle \tilde\www|\www\rangle_{\HH^{(c)}}
:=&\frac{1}{2}
\eta_{\bcdot\star}\eta_{\bcdot\star}\hspace{.2em}
\tilde\omega^{\hspace{.3em}\bcdots}_{\mu_1}\hspace{.2em}
\omega^{\hspace{.3em}\stars}_{\mu_2}\hspace{.2em}
dx^{\mu_1}\wedge dx^{\mu_2},\\
=&
\frac{1}{4}\epsilon_{\bcdott}\hspace{.2em}
\omega^{\hspace{.3em}\bcdots}_{\mu_1}\hspace{.2em}
\omega^{\hspace{.3em}\bcdots}_{\mu_2}\hspace{.2em}
dx^{\mu_1}\wedge dx^{\mu_2}.
\end{align*}
Under this bilinear form, the functional space $\varpi$ is considered a pseudo-Riemannian manifold with a metric tensor that is defined as
\begin{align*}
g^\omega_{\mu\nu}&:=\left[{\bm g}^\omega\hspace{-.1em}
\left(\tilde{\www},\www\right)\right]_{\mu\nu}=
\frac{1}{4}\epsilon_{\bcdott}\hspace{.1em}
\omega^{\hspace{.3em}\bcdots}_{\mu}\hspace{.1em}\omega^{\hspace{.3em}\bcdots}_{\nu}.
\end{align*}
A norm of the spin form $\|\www\|_{\HH^{(c)}}$ can be defined using the bilinear form as
\begin{align}
\left(\|\www\|_{\HH^{(c)}}\right)^{2}:=\int_{\Sigma_2}\langle\tilde\www|\www\rangle_{\HH^{(c)}}
=\int_{\Sigma_2}g^\omega_{\mu_1\mu_2}\hspace{.1em}
dx^{\mu_1}\wedge dx^{\mu_2}
\in\R,\label{blf}
\end{align}
where ${\Sigma_2}$ is the appropriate compact two-dimensional submanifold ${\Sigma_2}\subset\M$.
An element of the dual space $\tilde{\varpi}$ is provided by a linear combination of spin forms $\omega^{~ab}_{\mu}\in\varpi$; thus, functional spaces $\varpi$ and $\tilde{\varpi}$ are linearly equivalent.
Consequently, the Hilbert space can be obtained as $\HH^{(c)}\simeq\varpi$, and the Gel'fand triple becomes  $\varpi\simeq\tilde{\varpi}\simeq\HH^{(c)}$.
The same construction is possible for the surface form defining the standard bilinear form  as
\begin{align*}
\langle \tilde\SSS|\SSS\rangle_{\HH^{(c)}}
&:=\frac{1}{2}\tilde\SSS_\bcdots\wedge\SSS^\bcdots=
\left(3!\right)\hspace{.1em}\vvv.
\end{align*}

The abovementioned expressions of the commutation relations (\ref{QCCR}) are formal ones:
their exact representation is
\begin{align}
&\left[\widehat{\www}^{ab}(x),\widehat{\SSS}_{cd}(y)\right]:=
\left[\widehat{\omega}_{\mu}^{~~ab}(x),\widehat{\Varepsilon_{\nu}^{c}\Varepsilon_{\rho}^{d}}(y)\right],
\nonumber\\
&\hspace{5em}=-i\delta^{[a}_{c}\delta^{b]}_{d}
\epsilon_{\mu\nu\rho\sigma}
\delta^{(3)}(x^{\mu,\nu,\rho}-y^{\mu,\nu,\rho})
\int\delta^{(1)}(x^{\sigma}-y^{\sigma})dx^{\sigma}.\label{wS}
\end{align}
For a further discussion, the following simple $(1\hspace{-.1em}+\hspace{-.1em}3)$ coordinate-decomposition is considered. 
Suppose the global space-time manifold is filled with a congruence of geodesics whose tangent vector is time-like at any point on the line. 
A coordinate $x^0$ is obtained along the time-like vector on these geodesics. 
Using this coordinate system, the three-dimensional boundary is obtained as a manifold at $x^0=\tau$ (an \textit{equal-time} boundary).
By fixing $dx^\sigma=dx^0=d\tau$ (\textit{time} coordinate) in (\ref{wS}) instead of taking a sum concerning $\sigma$ and performing integration with respect to $d\tau$, one can obtain the standard \textit{equal-time} commutation relation.
Using this coordinate system, the three-dimensional boundary is obtained as a manifold at $x^0=\tau$ (an {\it equal-time} boundary) and the state is written as $|\Psi_\tau\rangle$.
Because spin connection $\omega^{~a}_{\mu~c}(x)$ is chosen as a general coordinate in the symmplectic manifold of general relativity, a state vector is a functional of spin connection\cite{Witten198846}, and the norm of state vector $\Psi_\tau$ is defined vy (\ref{blf}).
Although the existence of the norm is ensured, a negative norm state, $\langle\Psi_\tau | \Psi_\tau \rangle<0$, is included in $\HH^{(q)}$, it is eliminated from the physical state\cite{doi:10.1140/epjp/s13360-021-01463-3} owing to the Kugo--Ojima condition.

The quantum Hamiltonian operator corresponding to (\ref{CHG}) is obtained by replacing the surface form to corresponding operator (\ref{CCRSW1}) such that:
\begin{align}
\widehat{\HHH}_G&:=
-\frac{i}{2}
\vomega^{\star}_{~\bcdot}\wedge\vomega^{\bcdot\star}\wedge
\left[\frac{\delta~~}{\delta\www}\right]_{\star\star},\label{QHG}
\end{align}
and thus, the Schr\"{o}dinger equation can be represented as
\begin{align}
-\frac{i}{2}\vomega^{\star}_{~\bcdot}\wedge\vomega^{\bcdot\star}\wedge
\left[\frac{\delta~~}{\delta\www}\right]_{\star\star}|\Psi(\omega)\rangle&=E_G|\Psi(\omega)\rangle,
\label{SEeq}
\end{align}
where $|\Psi(\omega)\rangle\in\varpi$ is a state vector.
Equation (\ref{SEeq}) is referred to as the first-order Schr\"{o}dinger--Einstein equation (SE-I) in this study.
Here, operator ordering of ``$\widehat\www$-left and $\widehat\SSS$-right'' is exploited.
SE-I does not depend on an ordering convention owing to a relation
\begin{align*}
\frac{1}{2}\left[\frac{\delta~~}{\delta\www}\right]_\bcdots\left(\www^\bcdot_{~\star}\wedge\www^{\star\bcdot}\right)&=
\www^\bcdot_{~\bcdot}=0.
\end{align*} 
A formal solution of the SE-I is
\begin{align}
|\Psi(\omega)\rangle=&A\exp{\int\left(
\frac{i}{2\kh}\www^\bcdots\wedge\SSS_\bcdots
+\frac{i}{(\kh)^{3/2}}\lambda^\bcdot\VVV_\bcdot
\right)},\label{statev}\\
\VVV_a:=&\frac{1}{3!}\epsilon_{a\bcdots\bcdot}
\eee^\bcdot\wedge\eee^\bcdot\wedge\eee^\bcdot,\nonumber
\end{align}
where $A$ is an appropriate normalization constant and $\lambda^a$ are components of an arbitrary constant vector in $\TM$ such that ${\bm\lambda}=\lambda^\bcdot\partial_\bcdot$.
We fix a normalization constant such as 
\begin{align}
\left|\langle\Psi(\omega)|\Psi(\omega)\rangle\right|=1.\label{Anorm}
\end{align}
For the operator $\widehat\SSS=\delta/\delta\www$, the surface form, thus, vierbein form too, are treated as the independent function from the spin connection and yields 
\[
\frac{\delta\SSS}{\delta\www}=
\bm\epsilon\cdot\left(\frac{\delta\eee}{\delta\www}\wedge\eee\right)=0.
\]
The first term in the integral is expressed using the standard basis as
\begin{align}
\www^\bcdots\wedge\SSS_\bcdots&=
\frac{1}{2}\epsilon_{\bcdots\bcdots}\www^\bcdots\wedge\eee^\bcdot\wedge\eee^\bcdot,
\nonumber\\
&=\frac{1}{2}\epsilon_{\bcdots\bcdots}\hspace{.1em}
\omega_{\mu_1}^{~\bcdots}
\Varepsilon_{\mu_2}^\bcdot\hspace{.1em}
\Varepsilon_{\mu_3}^\bcdot\hspace{.1em} 
dx^{\mu_1}\wedge dx^{\mu_2}\wedge dx^{\mu_3}.\label{wSwS}
\end{align}
Whereas state vector $|\Psi(\omega)\rangle$ is a functional of the spin connection, integration in (\ref{statev}) is not functional integration but a Riemannian integration.
The state vector can be recognized as the mapping $|\Psi\rangle:\varpi\rightarrow\C$.
The second term of the integrand is a constant with respect to the functional of the spin connection; thus,it can be absorbed in the normalization constant.
When we choose  a point in $\varpi$, which corresponding to specify a function $\omega^{c}$, geometrical spin connection is given by expected value $\www^{(g)}:=\langle\Psi(\omega^{c})|\hat\www|\Psi(\omega^{c})\rangle$ of classical spin connection $\omega^{c}$. 

When the spin connection form is taken as the configuration variable and the surface form as the momentum variable,  a formal solution of the Einstein--Shr\"{o}dinger equation (\ref{SEeq}) is obtained obtained as in (\ref{statev}).
However, the operator representation (\ref{CCRSW1}) is not the unique possibility to realize the commutation relation (\ref{QCCR}). 
A set of operators such as
\begin{align}
\widehat{\www}^{ab}={i\kh}\left[\frac{\delta\hspace{1em}}{\delta\SSS}\right]^{ab}~~&\textrm{and}~~~
\widehat{\SSS}_{ab}=\SSS_{ab},
\label{CCRSW2}
\end{align}
are also possible.
The corresponding Schr\"{o}dinger--Einstein equation can be expressed as
\begin{align}
-\frac{\kh}{2}\eta_\start\SSS_\bcdots\wedge
\left[\frac{\delta\hspace{1em}}{\delta\SSS}\right]^{\bcdot\star}\wedge
\left[\frac{\delta\hspace{1em}}{\delta\SSS}\right]^{\star\bcdot}|\Psi\rangle
&=E_G|\Psi\rangle,\label{schrodingerA}
\end{align}
owing to the vacuum Hamiltonian (\ref{CHG}).
This equation is referred to as the second-order Schr\"{o}dinger--Einstein equation (SE-II).
Here, the operator ordering of ``$\widehat\www$-right and $\widehat\SSS$-left'' is employed.
SE-II is also independent of the operator-ordering convention, owing to the relation
\begin{align*}
\eta_\start
\left(\left[\frac{\delta~~}{\delta\SSS}\right]^{\bcdot\star}\SSS_\bcdots\right)
\wedge
\left[\frac{\delta\hspace{1em}}{\delta\SSS}\right]^{\star\bcdot}&=
\eta_\bcdots\left[\frac{\delta\hspace{1em}}{\delta\SSS}\right]^\bcdots=0.
\end{align*}
Direct calculations shows the solution (\ref{statev}) fulfils the equation (\ref{schrodingerA}).
As a consequence, representations SE-I and SE-II yield an equivalent solution.

The prequantum Hilbert space of the spin and surface forms is given by $\varpi=\sigma\subset L^2(\TM)$, where $L^2(\TM)$ represents the set of squared integrable functions in $\TM$.
The prequantum Hilbert space, denoted $\HH^{(q)}$, is a dual space of $\HH^{(c)}$ with respect to the functional derivative operator (\ref{CCRSW1}), and yields the Gel'fand triple $\HH^{(c)}\subset L^2(\TM)\subset \HH^{(c)*}:=\HH^{(q)}$.
Because the quantum operator $\widehat{\SSS}_{ab}$ includes imaginary unit $i:=\sqrt{-1}$, the dual space $\HH^{(q)}$ must be a set of  complex functions. 
Even though the spin connection is a square-integrable function, state vectors are not necessarily holomorphic functions.
Whereas the state vectors treated in this study are holomorphic, the analyticity of the state vectors must be confirmed in general.

\section{Application to Schwarzschild black holes}\label{ASBH}
The results obtained in \textbf{section \ref{QM}} are here applied to black hole solutions.
The quantization of black holes has a long history.
Bekenstein first discussed the mass spectrum of a Kerr black hole by analogy with a charged particle carrying spin\cite{Bekenstein1974}.
Louko and Winters-Hilt extended the discussion to Reissner--Nordstr$\ddot{\rm o}$m--anti-de\hspace{.1em}Sitter black holes using Hamiltonian thermodynamics\cite{PhysRevD.54.2647}.
Many authors have also considered an area spectrum of black holes, based on a quantum area-operator \cite{0264-9381-14-1A-006,0264-9381-15-6-001,0264-9381-18-22-310,0264-9381-20-9-305,0264-9381-25-20-205014}.
Other approaches have been proposed, derived from string theory\cite{Kiselev:2004vx} and loop quantum gravity\cite{PhysRevLett.110.211301}. 

The black-hole mass-spectrum based on SE-I is discussed and compared with that derived from the EBK condition in this section.
\subsection{Quantum Schwarzschild solutions}\label{QSS}
The Schwarzschild solution of the classical Einstein equation is given in polar coordinates $x^a=(t,r,\theta,\phi)$ as
\begin{align}
ds^2&=f^2(r)dt^2-\frac{dr^2}{f^{2}(r)}
-r^2\left(d\theta^2+\sin^2{\phi}\hspace{.1em}d\phi^2\right),\label{lmschw}
\end{align}
where $f^2(r)=1-2MG/r$ and $M$ is the black-hole mass measured by an asymptotic observer at an infinite distance away from the hole, at rest.
A vierbein form is provided from (\ref{lmschw}) as
\begin{align}
\eee^a_{\textrm{Schw}}&=\left(
fdt,f^{-1}dr,rd\theta,r\sin{\theta}\hspace{.1em}d\phi
\right).\label{virschw}
\end{align}
The above solutions and the torsion-less condition provide a unique representation of the spin form \cite{Kurihara:2016nap}:
\begin{align}
\vomega^{ab}=\left(
\begin{array}{cccc}
0&-GM/r^2dt& 0 & 0 \\
~& 0 &  fd\theta &
f\sin{\theta}\hspace{.1em}d\phi \\
~&~&0&~~\cos{\theta}\hspace{.1em}d\phi\\
~&~&~&0\\
\end{array}
\right).\label{www}
\end{align}
(The lower half of the matrix is omitted, given the antisymmetry of the spin form.)
The classical Hamiltonian (\ref{CHG}) can be expressed as
\begin{align}
\HHH_{\textrm{Schw}}=-\frac{1}{2\hbar\kappa}\vomega^{\bcdot}_{~\star}\wedge\vomega^{\star\bcdot}
\wedge\SSS_\bcdots,
=-\frac{1}{\hbar\kappa}\sin{\theta}\hspace{.1em}dt\wedge dr\wedge d\theta\wedge d\phi,\label{Hscw}
\end{align}
and, accordingly, a three-form object necessary to estimate the state vector is provided as
\begin{align}
\frac{1}{2\kh}\vomega^\bcdots\wedge\SSS_\bcdots=\frac{1}{\kh}\left(
\left(3GM-2r\right)\sin{\theta}\hspace{.1em}dt\wedge d\theta\wedge d\phi
+\cos{\theta}\hspace{.1em}dt\wedge dr\wedge d\phi
\right).\label{Lfm}
\end{align}
The expected value of the quantum Hamiltonian can be obtained using (\ref{Hscw})  as
\begin{align}
E_{\textrm{Schw}}(R)=\langle\Psi(\omega)|\widehat{\HHH}_{\textrm{Schw}}|\Psi(\omega)\rangle
=\frac{1}{\hbar\kappa}\int^{t_p}\int^R\int^{S^2}\HHH_{\textrm{Schw}}=-\frac{R}{l_p},\label{Eschw}
\end{align}
where $S^2$ is two-dimensional sphere and $|\Psi(\omega)\rangle$  is given by (\ref{statev}) with the normalization factor ${A}$ by (\ref{Anorm}).
A finite upper bound for the $r$-integration is set to $r=R$.
Here, (\ref{Eschw}) gives the expected value of the total gravitational energy in the two-dimensional sphere with a radius $R$ during the Planck time measured by the asymptotic observer.
The energy density per unit surface area at $r=R$ is provided as
\begin{align*}
\overline{E}_{\textrm{Schw}}(R)&=\frac{E_{\textrm{Schw}}}{4\pi (R/l_p)^2}=-\frac{1}{4\pi}\frac{1}{(R/l_p)}.
\end{align*}
Thus, the energy density at the event horizon of a black hole measured by the asymptotic observer becomes
\begin{align}
\Bigl|\overline{E}_{\textrm{Schw}}(R=2M G)\Bigr|&=\frac{1}{m_p}\left(\frac{\hbar}{8\pi GM}\right).
\label{BHTemp}
\end{align}
We note that the energy density of space time in a given region is a coordinate-dependent observable.
The energy density given here is an expected value in the asymptotic frame.
This energy density (in units of the Planck mass) is none other than the Hawking temperature\cite{hawking1975}.

The state vector (\ref{statev}) with form (\ref{Lfm}) is provided as
\begin{align}
|\Psi(r)\rangle&\propto
\exp{\left(4\pi i\left(2\frac{r}{l_p}-3\frac{M}{m_p}\right)\right)},\label{state2}
\end{align}
with integration over a unit sphere over the Planck time.
After the integration, the state vector is considered as a function of radial coordinate $r$.
The second term in (\ref{Lfm}) is related to the angular moment of the black hole, and is therefore ignored at present, because only non-rotating black holes are considered here.
The first observation is that the state vector is finite at the center of the black hole.
By analogy with a point mass in a quantum well, boundary conditions may induce a black-hole energy spectrum.
However, the precise form of these boundary conditions is still unknown, given our lack of understanding of quantum black holes.
The requirement that the quantum state in the black hole be a standing wave imposes a constraint on the phase difference between the center and the event horizon of the black hole:
\begin{align}
(2\pi i)\times n=\bigl|\log{\left(|\Psi(2GM)\rangle \right)}-\log{\left(|\Psi(0)\rangle\right)}\bigr|
=16\pi i \frac{M}{m_p},\label{bc12}
\end{align}
where $n\in\Z$.
This condition implies a minimum black-hole mass of $M_{min}=m_p/8$,
which can be satisfied by configuring the EBK quantization condition (\ref{Evac}) as
\begin{align}
M_n^\textrm{EBK}&=\frac{m_p}{2}\left(n+\frac{1}{4}\right),~~n=0,1,2,\cdots.\label{Mschw0}
\end{align}
This corresponds that setting $\nu=4$ because $[H^4(\M_5,\Z)]=1$ is obtained in Ref.\cite{Kurihara_2020}.

Mass spectra obtained in previous studies\cite{Bekenstein1974,0264-9381-20-9-305,0264-9381-18-22-310,Bekenstein:1997bt} seem to differ from our result.
Since, in previous studies, the black-hole mass spectrum was derived from an area spectrum (including effects due to charge and angular momentum), the spectrum condition depends on the square of the mass rather than on the mass itself. Thus, for a Schwarzschild black hole,
\begin{align}
\tilde{M}^2&=\frac{m^2_{p}}{2}\left(n+\tilde\mu\right),\label{BHEngy2}
\end{align}
where $\tilde\mu=0$ was proposed by Bekenstein\cite{Bekenstein1974,Bekenstein:1997bt}, Medbed\cite{0264-9381-25-20-205014}, and $\tilde\mu=1/2$ by Barvinsky, Das, Kunstatter\cite{0264-9381-18-22-310} and Gour, Medved\cite{0264-9381-20-9-305}.

\subsection{McVittie--Thakurta metric}\label{McVTha}
\subsubsection{Classical solution}
In section \ref{QSS}, a quantum effect of an isolated Schwarzschild black hole in the vacuum is examined.
The present section considers a black hole on a scalar field background.
In 1933, McVittie reported an exact solution to the Einstein equation, representing the Schwarzschild black hole in an expanding universe\cite{1933MNRAS..93..325M}. In 1966, he again discussed a solution of a Schwarzschild black hole on the FLWR background metric\cite{1966ApJ...143..682M}.
A McVittie metric can be interpreted as a solution of the Schwarzschild black hole with a time-dependent mass.
In 1981, Thakurta extended the scope of the McVittie metric to rotating black holes\cite{1981InJPh..55..304T}.
(Yet another solution for back holes in an expanding universe was found by Gibbons and Maeda\cite{PhysRevLett.104.131101}.)
The McVittie and Thakurta solutions have given rise to many cosmological studies \cite{PhysRevD.58.064006,0264-9381-16-4-012,0264-9381-16-10-310,Sultana2005,Firouzjaee:2008gs,PhysRevD.81.104044,Bolejko:2011jc,PhysRevD.84.044045,PhysRevD.86.124020,PhysRevD.87.064030,PhysRevD.85.083526,Barvinsky:2000gf,PhysRevD.91.084043,PhysRevD.95.084031}. 
Thakurta's representation of a metric with a vanishing angular momentum is referred to as the McVittie--Thakurta metric in this study.
The McVittie--Thakurta metric is suitable for analyzing a interplay between a black hole and a scalar field.

Consider a system consisting of the Schwarzschild black hole and a classical scalar field.
The McVittie--Thakurta metric in this case can be expressed\cite{PhysRevD.95.084031} similarly to (\ref{virschw}) as
\begin{align*}
\eee_{\rm{McT}}^a&=\left(
f(r)dt,\hspace{.1em}f^{-1}(r)\Omega(t)dr,
\hspace{.1em}\Omega(t)r\hspace{.1em}d\theta,
\hspace{.1em}\Omega(t)r\sin{\theta}\hspace{.1em}d\phi
\right),
\end{align*}
where $\Omega(t)$ is a scale function induced by the existence of the scalar field.
We note that physical time $t$ is used in this study instead of utilizing cosmological time $d\tau=\Omega(t)dt$ as in Ref.\cite{PhysRevD.95.084031}.
According to Ref.\cite{PhysRevD.95.084031}, the cosmological mass $M_\Omega$ and radius $R_\Omega$ are introduced, respectively, as
\begin{align*}
M_\Omega(t)=M\Omega(t)~~&\textrm{and}~~ R_\Omega(t)=r~\Omega(t),
\end{align*}
and the corresponding Friedmann equations are obtained as
\begin{align}
\Gh\frac{3H^2}{f_\Omega^2}&=~~8\pi G\hspace{.2em}T^{00},\label{sFME1}\\
\Gh\frac{3H^2+2\dot{H}}{f_\Omega^2}&=-8\pi G\hspace{.2em}T^{11},\label{sFME2}\\
G^2\hbar\frac{M_\Omega H}{f_\Omega^2R_\Omega^2}&=-4\pi G\hspace{.2em}T^{10},\label{sFME3}
\end{align}
with
\begin{align*}
f_\Omega^2&=1-\frac{2GM_\Omega}{R_\Omega}=1-\frac{2GM}{r},
\end{align*}
where $H=\partial_t\Omega/\Omega$ is the Hubble ``constant''.
In this study, the energy-momentum tensor $T^{ab}$ is induced by classical scalar field $\varphi$.
A geometrical representation of the scalar field exploited in this analysis is summarized in Appendix~\ref{ap1}. 
This analysis defines the energy-momentum tensor $T^{ab}$ such that $G\hspace{.2em}T^{ab}$ is dimensionless.
The scalar field $\varphi$ and its potential energy $V(\varphi)$ have physical dimensions of $(mass)$ and $(mass)^4$, respectively, as in a standard definition.
The energy-momentum tensor is provided owing to the scalar field as
\begin{align*}
T^{00}/G&=
\frac{1}{4}(\partial_t{\varphi})^2+\frac{1}{4}(\partial_r{\varphi})^2
+\frac{1}{2\hbar^2}V(\varphi),\\
T^{11}/G&=
\frac{1}{4}(\partial_t{\varphi})^2+\frac{1}{4}(\partial_r{\varphi})^2
-\frac{1}{2\hbar^2}V(\varphi),\\
T^{10}/G&=-\frac{1}{2}(\partial_t{\varphi})(\partial_r{\varphi}).
\end{align*}
Here, the scalar field is assumed to be isotropic but neither uniform nor static. It is therefore a function of $t$ and $r$, and denoted $\varphi=\varphi(t,r)$.
The equation of motion for the scalar field under this condition can be expressed as
\begin{align}
\hbar^2\left[
\frac{1}{f_\Omega^2}\left(
\partial_t^2\varphi+3H\partial_t\varphi
\right)
-\frac{f_\Omega^2}{\Omega^2(t)}\partial_r^2\varphi
-2\frac{1-GM_\Omega/R_\Omega}{R_\Omega\Omega(t)}\partial_r\varphi
\right]+\partial_\varphi V(\varphi)=0.\label{KGMT}
\end{align}
Under slow rolling and almost flat conditions for the scalar field, it is assumed to be a linear function  with respect to both $t$ and $r$  such that
\begin{align}
\varphi(t,r)&=\sqrt{\frac{\hbar}{G}}~\varphi_0+\frac{\varphi_t}{G}t +\frac{\varphi_r}{G}r,\label{phiexp}
\end{align}
with $\varphi_0\gg\varphi_t,\varphi_r$.
We note that all constants $\varphi_\bullet$ are set to dimensionless.
In addition, the potential energy $V(\varphi)$ is required to satisfy $\mathcal{O}(V/m_p^4)\geq\mathcal{O}\left(\varphi_0^2/m_p^2\right)$.
Using these approximations, the equation of motion (\ref{KGMT}) is expressed as
\begin{align}
\frac{\hbar^2}{G}\left(
3\frac{H}{f_\Omega^2}\varphi_t
-2\frac{1-GM_\Omega/R_\Omega}{R_\Omega\Omega(t)}\varphi_r
\right)
+\partial_\varphi V(\varphi)&=0.\label{KGMT2}
\end{align}
Solutions of the Friedmann equations (\ref{sFME1}) and (\ref{sFME2}) are then provided as
\begin{align}
H&=\pm\frac{f_\Omega}{\hbar}\sqrt{\frac{\kappa V(\varphi)}{3\hbar}}\label{HCsol},\\
\Omega(t)&=\exp{(Ht)},\label{HCsol2}
\end{align}
with the initial condition $\Omega(t=0)=1$.
Although no concrete functional form of the potential energy $V(\varphi)$ is introduced here, the solution (\ref{HCsol}) is consistent with the equation of motion (\ref{KGMT2}) within above approximations. 
From this solution, the energy flux across a two-dimensional sphere with radius $R_\Omega$ during unit time is given by
\begin{align}
GT_{01}&=\pm\frac{G}{\hbar}\sqrt{\frac{V(\varphi)}{12\pi}}
\frac{M_\Omega/m_p}{f_\Omega\left (R_\Omega/lp\right)^2}
=\frac{1}{2}\varphi_t\varphi_r.\label{GT01}
\end{align}
Above solutions are still meaningful in the black hole if we assume $V(\varphi)<0$ in the region specified by $r<2GM$. This assumption is reasonable because the sign of $g_{00}$ is negative in the black hole, in contrast with that in the range $r>2GM$.
The Hubble ``constant'' can have both positive and negative solutions, as shown in (\ref{HCsol}).
The time-dependent mass $M_\Omega$ and the radial length $R_\Omega$ are scaled equally so that $\dot{M}_\Omega/{M}_\Omega=\dot{R}_\Omega/{R}_\Omega=H$, as in Ref.\cite{PhysRevD.95.084031}.
Actual observations in the expanding universe are discussed in refs.\cite{1971ApJ...168....1N,Cooperstock:1998ny,doi:10.1119/1.3699245}.
In the present case, the diameter of the event horizon increases with time for positive solutions ($H>0$). The expansion of the event horizon decreases the surface temperature, as shown in (\ref{BHTemp}) 
as a result of the negative energy flow of (\ref{GT01}). This signifies that the scalar-field energy is absorbed by the black hole.
On the other hand, the solution with negative $H$ yields energy emission from the black hole through the positive solution of  (\ref{GT01}).
We note that the sign of the energy flow is positive when the black-hole radius increases.
Absorption or emission of the scalar energy may occur spontaneously with a probability that is calculated using the quantum-gravity theory proposed here.

\subsubsection{Quantum correction}
The Hamiltonian of the Schwarzschild metric is considered as a Born Hamiltonian that includes  a scalar field as a small perturbation. 
The gravitational effect of a scalar field in the Lagrangian form is expressed as (\ref{LagrangianCF2}), as given in Appendix~\ref{ap1}, and is proportional to the volume form.
The following approximation is used to investigate the possible interplay between black holes and the scalar field:
The mass spectrum is provided via the Born approximation as (\ref{Mschw0}).
Accordingly, state vectors are parameterized using the mass eigenstates.
The transition probability from one mass state to another is estimated using Fermi's golden rule with the interaction Hamiltonian.
 
Schwarzschild Hamiltonian $\HHH_{\textrm{Schw}}$ given in (\ref{Hscw}) is denoted $\HHH_0$ hereafter.
The Schwarzschild Hamiltonian at the Born level $\HHH_0$ and its eigenstate $|\Psi_0\rangle$ with a black-hole mass $M_n$ can be expressed as
\begin{align}
\HHH_0&=-\frac{1}{4\pi\kh}\frac{1}{r^{2}}~\vvv,\label{BHF}\\
|\Psi_0\rangle&=|r;n\rangle=\exp{\left(
4\pi i\left(
\frac{2r}{l_p}-\frac{3M_n}{m_p}\right)
\right)},
\end{align}
where $M_n$ is given by (\ref{Mschw0}).
The scalar Hamiltonian density $\HH$ is introduced from a Hamiltonian four-form as $\HHH=\HH~\vvv/(\kh)^2$.
From the Born Hamiltonian-form (\ref{BHF}), the Born Hamiltonian density $\HH_0$ is provided as\begin{align}
\HH_0&=-\frac{1}{\left(r/l_p\right)^2}.
\end{align}
The McVittie--Thakurta Hamiltonian-form is provided as
\begin{align}
\HHH_{\textrm{McT}}&=-\frac{1}{4\pi\kh}\left(
\frac{1}{R_\Omega^{2}}+3\left(\frac{H}{f_\Omega}\right)^2
\right)\vvv_{\textrm{McT}},\label{HMcT}
\end{align}
where $\vvv_{\textrm{McT}}$ is the four-dimensional volume form of the McVittie--Thakurta metric, that is represented as
\begin{align*}
\vvv_{\textrm{McT}}=
\Omega^3(t)r^2\sin{\theta}~
dt\wedge dr\wedge d\theta\wedge d\phi
=dt\wedge(R_\Omega dR_\Omega)\wedge(R_\Omega\sin{\theta}d\theta)\wedge d\phi.
\end{align*}
From the Hamiltonian four-form (\ref{HMcT}), the Hamiltonian density $\HH_{\textrm{McT}}$ is obtained as
\begin{align*}
\HH_{\textrm{McT}}=-\frac{1}{4\pi\kh}\left(
R_\Omega^{-2}+3\left(\frac{H}{f_\Omega}\right)^2
\right)
=\HH_0\exp{(-2Ht)}-\frac{3}{4\pi\kh}\left(\frac{H}{f_\Omega}\right)^2,
\end{align*}
where solution (\ref{HCsol2}) is used.
The difference between the two Hamiltonian densities is defined as $\delta\HH:=\HH_{\textrm{McT}}-\HH_0$.

\begin{figure*}[tb]
 \begin{center} 
\includegraphics[width={\linewidth}]{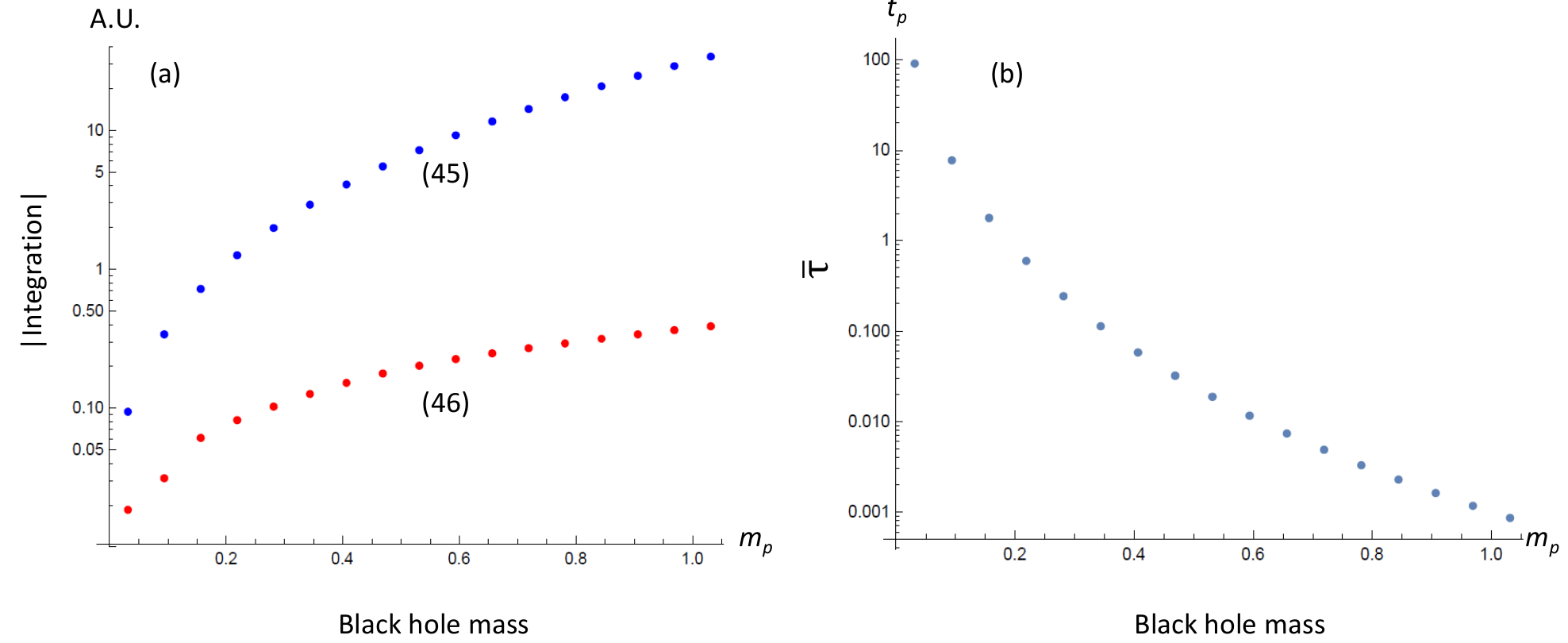}%
 \caption{\footnotesize
The transition rate (\ref{TM0}) is calculated for the potential energy $V(\varphi_0)=1/m_p^4$ in Planck units. 
Integration is performed within the event horizon over time $t_p$.
Numerical results of (\ref{transition}) and (\ref {rint}) with black-hole masses ranging from the minimum mass ($m_p/32$) to the Planck mass are shown in (a).
The mean life time for spontaneous emission of the scalar field is shown in (b) with the same parameters as those in (a).
} 
\label{fig1}
 \end{center} 
\end{figure*}

The results of quantum general-relativity given in section \ref{McVTha} are applied to this system to maintain the  classical character of the scalar field. 
The transition rate from the initial state $|\Psi_0^i\rangle$ to the final state $|\Psi_0^f\rangle$ (denoted $\Gamma_{i\rightarrow f}$) is estimated owing to the Fermi's golden rule:
\begin{align}
\Gamma_{i\rightarrow f}&=\left|
\langle \Psi_0^f|\delta\HH|\Psi_0^i\rangle
\right|^2
\delta\left(
E_0^f-E_0^i\pm E_\varphi
\right),\label{TM0}
\end{align}
where $E_0^{i,f}$ are the energy eigenvalues corresponding to the respective states $|\Psi_0^{i,f}\rangle$ and $E_\varphi$ is the energy absorbed or emitted as a result of the scalar field.
Transitions from one mass state to another are possible by virtue of the energy flow between the black hole and the scalar field given in (\ref{GT01}).
The transition matrix can be calculated as
\begin{align}
\langle \Psi_0^f|\delta\HH|\Psi_0^i\rangle=&\int\left(
\HH_0\exp{(-2H(r)t)}-\HH_0-\frac{3}{4\pi\kh}\left(\frac{H(r)}{f_\Omega}\right)^2
\right)\nonumber\\
&\times\langle r;n|r;n\pm 1\rangle~
\vvv,\label{TM}
\end{align}
where $H(r)$ is a solution of the Friedmann equation in (\ref{HCsol}).
Here, the potential energy of the scalar field is treated as a constant exploiting an approximation (\ref{phiexp}), and Thus, $H(r)$ is a function of $r$ only, via the function $f_\Omega$.
For on-shell black holes, a squared state vector is normalized for any $0\leq n\in\Z$ as
\begin{align}
|\Psi_0|^2:=
\langle r;n|r;n\pm 1\rangle=1.
\end{align}
The second and third terms in the integration (\ref{TM}) can be integrated straightforwardly as
\begin{align}
-\frac{1}{4\kh}\int\left(
\frac{1}{r^2}+3\left(\frac{H(r)}{f_\Omega}\right)^2\right)|\Psi_0|^2~\vvv_0=
{N_n}
\left(1-
\frac{4\pi N_n^2}{3}
\frac{V(\varphi_0)}{m_p^4}
\right)\frac{\tau}{t_p},\label{transition}
\end{align}
where $N_n=(n+1/4)$.
The integration is performed within the event horizon, up to time $\tau$.
Note that the potential energy $V$ has physical dimensions of $(mass)^4$, and consequently, the integration (\ref{transition}) is dimensionless.
Integrations of the first term of (\ref{TM}) can be performed except for a radial coordinate as
\begin{align}
-\frac{1}{4\kh}\int\frac{\exp{\left(-2H(r)t\right)}}{r^2}|\Psi_0|^2\vvv=
\frac{1}{2\kh}\int_0^{2G M_{n}}\frac{\exp{\left(-2H(r)\tau\right)}-1}{H(r)}dr.\label{rint}
\end{align}
In summary, the transition matrix can be expressed as $\langle \Psi_0^f|\delta\HH|\Psi_0^i\rangle=(\ref{transition})+(\ref{rint})$.
Despite not being expressed using elementary functions, the integral converges.
Integration results with negative $H(\tau)$ values are simply the complex conjugates of those with positive $H(\tau)$ values, and therefore the absolute value is independent of the choice of a sign of $H(\tau)$. 
The results of numerical integrations of (\ref{rint}) with $V=m_p^4$ and $\tau=t_p$ are shown in  Figure~\ref{fig1}-(a) together with the numerical values of (\ref{transition}).
These results can be interpreted as signifying that $\Gamma_{i\rightarrow f}$, given by (\ref{TM0}), is the probability of a transition occurring from a state with mass $M_{n}$ to one with $M_{n\pm1}$ over a period $\tau$.
Because the contribution from (\ref{transition}) is dominated in the transition probability as shown in Figure~\ref{fig1}-(a), the transition probability is almost proportional to the time period.
Therefore, the probability for $N$ transitions occurring per unit time, denoted $p(N)$, should have a Poisson distribution $p(N)=\Gamma^N\exp{(-\Gamma)}/N!$, while the mean time interval between two successive transitions should have an  exponential distribution with mean value $\overline{\tau}=1/\Gamma$.
Numerical results of the mean interval as a function of the hole mass near the ground state with $V(\varphi_0)=m_p^4$ are shown in Figure~\ref{fig1}-(b).
The transition from one state to another may occur spontaneously, in an analogous manner to spontaneous photoemission from a hydrogen atom.

\subsubsection{Comparison with Hawking radiation}
The energy emission from a black hole discussed in this study resembles  Hawking radiation\cite{hawking1975}, but this is not true in actual fact.
Hawking radiation, like the Unruh effect\cite{PhysRevD.14.870}, is induced by a kinetic term of the scalar field through the metric tensor as $g^{\mu\nu}\partial_\mu\varphi\partial_\nu\varphi$.
On the other hand, the radiation treated in this study arises from the potential energy of the scalar field, as shown in (\ref{GT01}).
Moreover, whereas the intensity of Hawking radiation is proportional to the reciprocal black-hole mass, it is proportional to the black-hole mass itself in our case.
The mean lifetime of a black hole with a mass $M$ due to Hawking radiation is provided as $\overline{\tau}_H(M)=5120\pi G^2M^3/\hbar$, that gives a lifetime for the black hole with the Planck mass as $\overline{\tau}_H(m_p)\sim10^5t_p$. 
On the other hand, a mean lifetime of a black hole due to quantum gravity with the scalar field $V(\varphi=0)=m_p^4$ provides a much shorter life time as $\overline{\tau}(m_p)\sim10^{-3}t_p$. 
A contribution of the Hawking radiation to a mean lifetime is negligibly small compared with that due to the quantum gravitational effect near the ground state.

A cosmological black hole in the real universe (e.g., with mass $M=2M_{\odot}$) have a mean life-time of  $\overline{\tau}_H(2M_{\odot})\sim6.0\times10^{123}t_p$ as a result of Hawking radiation, where the solar mass is $M_{\odot}\sim3.6\times10^{39}m_p$.
Cosmological black holes have a longer lifetime than the age of the universe.
In the case of quantum gravity, however, the mean interval is proportional to $\overline{\tau}\propto[(M/m_p)^{3} V(\varphi_0)/m_p^4]^{-1}$ from (\ref{transition}).
In this case, the mean time-interval due to the quantum gravitational effect becomes very short, and emission and absorption are completely balanced.
As a result, a cosmological black hole becomes static.
In conclusion, our model of quantum gravity does not yield any observable effect for cosmological black holes arising from the interaction between the black holes and a scalar field in the real universe.

\section{Cosmological constant problem}\label{CCP}
The previous section discusses The impact of quantum effects on black-hole dynamics in a background owing to the scalar field.
This section treats the interplay between gravity and the vacuum energy of a scalar field in terms of the statistical mechanics of black holes, based on the mass spectrum (\ref{Mschw0}) given in Section \ref{QSS}.
For simplicity, we assume that the universe is filled with a scalar field only, and the quantum scalar field has vacuum energy that interacts with space-time.
This point was first discussed by Zel'dovich \cite{Zeldovich:1967gd} in 1969, where a significant discrepancy was noted between the expected and observed values of the vacuum energy.
Weinberg estimated the vacuum-energy density induced by the ground-state energy of the scalar-field\cite{RevModPhys.61.1}:
\begin{align}
\langle\rho_{vac}\rangle&=\int_0^{\lambda_G}\frac{4\pi k_s^2dk_s}{(2\pi)^3\hbar^3}
\frac{1}{2}\sqrt{k_s^2+m_s^2}\simeq
\frac{\lambda_G^4}{16\pi^2\hbar^3},\label{vaceng}
\end{align}
where $k_s$ and $m_s$ are the momentum and mass of the scalar filed, respectively.
If the energy cut-off $\lambda_G\gg m_s$ of the integration is set to the Planck energy, the vacuum-energy density becomes $\langle\rho_{vac}\rangle\simeq6.3\hbar^{-1}G^{-2}\simeq1.4\times10^{74}~\textrm{GeV}^4$.
Here the physical dimensions of $\langle\rho_{vac}\rangle$ are $(mass)/(length)^{-3}$ in our units and is $(energy)^4$ in natural units.
This value is considerably greater than the observed value of dark energy $\rhpDE\simeq10^{-47}~\textrm{GeV}^4$\cite{Ade:2015xua,Olive:2016xmw}.
This discrepancy constitutes the so-called ``cosmological constant problem'' and is one of the most notable discrepancies between an established theory and a precise observation.
Although several explanations have been put forward, there is still no consensus.
Recent reviews can be found in references\cite{Nobbenhuis:2004wn, Martin:2012bt, Burgess:2013ara, Padilla:2015aaa}.

As shown in {\bf section \ref{ASBH}}, the vacuum energy of a scalar field encompassing the universe leads to the formation of a black hole that should contribute to decelerating the expansion of the universe.
Thus, all of the vacuum energy estimated above does not simply cause the universe's expansion.
The net effect of the expansion must be determined by considering the balance between two opposing effects resulting from the scalar field and black holes.
The Friedmann equations (\ref{sFME1}) and (\ref{sFME2}) are here used to analyze the cosmological constant problem.
The analysis in section \ref{McVTha} considered only the contribution from the scalar field to the energy-momentum tensor.
The potential energy of the scalar field contributes to the Hubble constant through (\ref{HCsol}) with $f_\Omega\rightarrow1$.
It is not enough to estimate the contribution of a single black hole.
The analysis must treat the energy-momentum tensor in the case of multiple black holes.
The solution for a single Schwarzschild black hole on an FLWR background is represented using the McVittie--Thakurta metric.
However, a solution of multiple holes on the same background is not the exact solution owing to the non-linearity of the Einstein equation.
A further assumption is therefore required to treat multiple black holes.
The distance between separate black holes makes their interaction energy much smaller than the energies arising from black-hole masses and the scalar field, making a linear approximation sufficient for discussing their contribution to the universe's expansion.

Although black holes can be treated as perfect fluids under the above assumption, their behavior differs from cosmic dust in the universe.
The energy-momentum tensor for a perfect static fluid in an isotropic and homogeneous universe can be simply written using the density $\rho$ and pressure $p$ of the fluid as $T^{00}=\hbar G\rho$ and $T^{11}=T^{22}=T^{33}=\hbar Gp$, respectively.
The state equation for cosmic dust is $p_d=0$, and the Friedmann equation yields the solution $\rho_d\Omega(t)^3=C$, where $p_d$ and $\rho_d$ are, respectively, the pressure and density of the dust and $C$ is a constant determined by the boundary conditions.
Because the total three-dimensional volume of the universe is proportional to $\Omega(t)^3$, this solution shows expressing the conservation of the total amount of dust.
On the other hand, our analysis's total number of black holes is not conserved because this number depends on the balance with the scalar field.
Black holes interact with each other through gravity, even though the interaction energy is assumed to be much smaller than their mass.
The local density variation of black holes should propagate in the form of gravitational wave\cite{PhysRevLett.116.241103,2016PhRvL.116x1103A,TheLIGOScientific:2017qsa} at the speed of light, i.e., $dp_{\hspace{-.1em}B\hspace{-.1em}H}/d\rho_{\hspace{-.1em}B\hspace{-.1em}H}=c^2$, where $p_{\hspace{-.1em}B\hspace{-.1em}H}$ and $\rho_{\hspace{-.1em}B\hspace{-.1em}H}$ are the pressure and density of the black holes, respectively.
Therefore, the state equation for black holes is $p_{\hspace{-.1em}B\hspace{-.1em}H}=c^2\rho_{\hspace{-.1em}B\hspace{-.1em}H}$.
The speed of light $c$ is dimensionless and has a numerical value $c=1$ in our units and will be omitted from formulae.
A perfect fluid with state equation $p=\rho$, henceforth referred to as an ultra-stiff fluid, was first discussed by Zel'dovich\cite{Zeldovich1961}.
Many authors have discussed cosmology in terms of such an ultra-stiff fluid \cite{MarkHeinzle2012,Chavanis:2014lra,Galiakhmetov2016}.
On the other hand, the state equation of the scalar field is given by $p_{s}=-\rho_{s}$, where $p_{s}$ and $\rho_{s}$ are the pressure and density of the scalar field, respectively.
The conservation of the total energy $(\rho_{\hspace{-.1em}B\hspace{-.1em}H}+\rho_{s})\Omega(t)^3=C$ is assured when the total pressure is zero, i.e., $p_{\hspace{-.1em}B\hspace{-.1em}H}+p_{s}=0$, which is equivalent to $\rho_{\hspace{-.1em}B\hspace{-.1em}H}=\rho_{s}$.
In other words, if the energy densities of black holes and the scalar field are precisely equal, the vacuum energy of the scalar field does not contribute to the expansion of the universe because black holes introduce a counter pressure.

Although the cosmological constant is tiny compared with the vacuum energy density, it is not precisely zero.
The cosmological constant problem is thus formulated as the question: {\it ``Why is the vacuum energy density owing to the scalar field not completely balanced by the black hole density?''}.
We treat this issue via a statistical-mechanical approach.
The universe is then treated as an isolated system, including a scalar field and black holes, referred to as a hole-field system.
The total volume of the universe and the number of black holes contained in it are not constant.
Because, as discussed above, the net pressure is assumed to be zero. 
The hole-field system does not perform any mechanical work on the universe during its expansion, and thus the total energy is conserved.
The reciprocal temperature of the scalar field is given by $\beta_{s}=(\hbar G)^{-3/2}\langle\rho_s\rangle^{-1}$ in our units, whose physical dimensions is $(mass)^{-1}$.
The vacuum energy (\ref{vaceng}), with an energy cut-off at the Planck energy, is exploited for quantitative analysis.
Assume the hole-field system is at thermal equilibrium, the reciprocal temperature of the black holes is the same as that of the scalar field, i.e., $\beta_{\footnotesize BT}=\beta_{s}=16\pi^2m_p^{-1}$.
In reality, the universe cannot be considered naively to be at thermal equilibrium because it is not causally connected.
This problem will be discussed later in this section.

A black hole mass-spectrum has a constant mass gap, and thus, a hole in the $n$th exited state is counted as $n$ grand-state holes.
One black hole with mass $M_i$ is statistically equivalent to $i$ black holes at the same location.
Because the number of black holes in excited states is not fixed, a number density of holes should be treated using the grand canonical ensemble and Bose statistics.
Whereas black-hole masses vary on the order of the Planck mass, that of the scalar field is naturally expected to be on the electroweak scale (i.e., $10^{2}$ GeV) or the grand unification scale ($10^{15}$ GeV).
Therefore, the level density of the scalar field is much denser than that of black holes, which justifies treating the scalar field as a heat bath\cite{feynman1998statistical}.
The average number of black holes of mass $M_n$ can be estimated as
\begin{align}
\langle N^n_{\hspace{-.2em}B\hspace{-.1em}H}\rangle&=
\frac{1}{e^{\beta_{\hspace{-.1em}B\hspace{-.1em}H}(M_n-\mu_{\hspace{-.1em}B\hspace{-.1em}H})}-1},\label{Ndecity}
\end{align}
and the black hole energy density in the universe is given by
\begin{align}
\langle \rho_{\hspace{-.1em}B\hspace{-.1em}H}\rangle l_p^3&=
\sum_{n=1}^\infty\frac{M_n}{e^{\beta_{\hspace{-.1em}B\hspace{-.1em}H}
(M_n-\mu_{\hspace{-.1em}B\hspace{-.1em}H})}-1},
\label{engdecity}
\end{align}
where $\mu_{\hspace{-.1em}B\hspace{-.1em}H}$ is the chemical potential.
The black hole mass is estimated using (\ref{Mschw0}).
The chemical potential can be determined by requiring $\langle \rho_{\hspace{-.1em}B\hspace{-.1em}H}\rangle=\rho_{s}$ in (\ref{engdecity}).
When the energy cut-off is set to $m_p$, the chemical potential is determined numerically to be $\mu=2.3\times10^{-1}m_p$.
This value ensures that the average value $\langle N^n_{\hspace{-.2em}B\hspace{-.1em}H}\rangle$ is positive for any $n\geq 0$.
The number of black holes in the real universe can be estimated using these assumptions.
For example, the total number of black holes in unit volume $l_p^3$ can be calculated numerically using (\ref{Ndecity}) as $N_{\hspace{-.2em}B\hspace{-.1em}H}=\sum_{n=0}\langle N^n_{\hspace{-.2em}B\hspace{-.1em}H}\rangle=2.5\times10^{-2}$.

As mentioned above, it is na\"{i}ve to assume that the real universe as a whole is at thermal equilibrium with the hole-field system.
Cosmic inflation is the most realistic hypothesis for realizing thermal equilibrium in the whole universe \cite{Starobinsky:1979ty,Starobinsky:1980te,Mukhanov:1981xt,1981MNRAS.195..467S,PhysRevD.23.347,Linde1982389,PhysRevLett.48.1220,Bezrukov2008703,Bezrukov2014249,Barvinsky:2008ia,BenDayan:2010yz,1475-7516-2012-02-008}.
This scenario posits that some scalar field caused an exponentially rapid expansion of the universe at its very beginning, induced by the negative pressure of the scalar field.
The complete balancing of this negative pressure with the positive pressure from black holes would have precluded cosmic inflation.
A possible solution is that the early universe was small enough to be at thermal equilibrium with the hole-scalar system, with a net zero average pressure.
Subsequently, statistical fluctuations gave rise to large negative pressures in the universe, followed by an exponentially rapid expansion (cosmic inflation).
During this inflation period, the number of black holes also increased exponentially because their number density was conserved, as shown above.
Cosmic inflation finished when the net pressure returned to zero as a result of the suppression of statistical fluctuations with the increasing number of black holes: $\delta N_{\hspace{-.2em}B\hspace{-.1em}H}/N_{\hspace{-.2em}B\hspace{-.1em}H}\sim N_{\hspace{-.2em}B\hspace{-.1em}H}^{-1/2}$.
A detailed quantitative justification of this hypothesis is beyond the scope of the present study.
For the time being, the following discussions assume that the universe was at thermal equilibrium after cosmic inflation.

By the central limit theorem, a statistical fluctuation of the system with $N$ samplings is given as $\delta \hspace{-.1em}N=N^{1/2}$.
The total number of background black holes in the universe is given by $N_{\rm{tot}}=4\pi r_{*}^3N_{\hspace{-.2em}B\hspace{-.1em}H}/3$, where $r_{*}=2.7\times10^{61}\hspace{.1em}l_{p}$ is the comoving radius of the observable universe.
We note that the number of samplings that contribute to statistical fluctuations is not simply the number of black holes in the current universe and must be determined by considering the universe's entire history.
A clear upper bound may be the total number of black holes in the whole four-dimensional volume of the universe at present, which can be estimated as
\begin{align}
N_{\hspace{-.1em}\rm{max}}&=\frac{4\pi}{3} r_{*}^3\hspace{.1em}t_{0}N_{\hspace{-.2em}B\hspace{-.1em}H}\sim3.4\times10^{244},
\end{align}
where $t_{0}=8.1\times10^{60}~l_p$ is the age of the universe.
Therefore, the statistical fluctuation in the total number of black holes due to the central limit theorem is $\delta N_{\hspace{-.1em}\rm{max}}=1/\sqrt{N_{\hspace{-.1em}\rm{max}}}\sim5.5\times10^{-123}$.
This causes a fluctuation in the net vacuum energy as
\begin{align}
\delta\rho_{\rm{vac}}&=\rho_{\rm{vac}}\hspace{.1em}\delta\hspace{-.1em}N_{\hspace{-.1em}\rm{max}}\sim 2.2\times10^{-48}~\textrm{GeV}^4,
\end{align}
where the vacuum energy of $\rho_{\rm{vac}}=1.4\times10^{74}~\textrm{GeV}^4$ is exploited.
This specifies a lower bound to the vacuum energy fluctuations, consistent with the measured value of the dark energy $\rhpDE\simeq2.6\times10^{-47}~\textrm{GeV}^4$.
Possible reasons for the discrepancy are the overestimation of the total number of black holes and the existence of other sources of fluctuation.
However, in this discussion, the central limit theorem results in a cancellation of more than a hundred orders of magnitude, making a fine-tuning unnecessary.
\section{Summary}\label{SUMMARY}
This study applies quantum general relativity to a Schwarzschild black hole and a hole-field system.
A quantum equation of gravity is proposed by considering the geometric quantization of general relativity.
Consequently, the expected value of the quantum Hamiltonian for a Schwarzschild black hole is consistent with the semi-classical Hawking temperature at the event horizon.
With several assumptions, the mass spectrum of Schwarzschild black holes is obtained using the EBK quantization condition.
A thermodynamic consideration of the hole-scalar-field system suggests that the universe is in thermal equilibrium of black holes and the ground-state energy of the scalar field.
By applying the grand canonical ensemble to black holes interacting with the vacuum energy of the scalar field, we can understand the measured dark energy in terms of statistical fluctuations in the number of black holes.
 
\section*{Acknowledgements}
The kind hospitality of all members of the theory group of Nikhef, particularly Prof. J. Vermaseren and Prof. E. Laenen, is gratefully acknowledged.
A significant part of this study was conducted during my stay at Nikhef in 2016 and 2017.
In addition, I would like to thank Dr.~Y.~Sugiyama and Prof. Fujimoto for their continued encouragement and fruitful discussions.
\begin{appendix}
\section{Geometrical representation of the scalar field}\label{ap1}
The Lagrangian of the scalar field $\varphi(x)$ can be expressed in the vierbein formalism as
\begin{align}
\LLL_S
&=\frac{1}{4!}\SSS_\bcdots\wedge\left[
\eta^\start\iota_\star\sss^\bcdot\wedge\iota_\star\sss^\bcdot
-V(\varphi)\eee^\bcdot\wedge\eee^\bcdot\right],
\label{LagrangianCF}
\end{align}
where $V(\varphi)$ is the potential energy and $\iota_a=\iota_{\xi^a}$ is the contraction with respect to the local coordinate-vector field $\xi^a=\eta^{a\bcdot}\Varepsilon^\mu_\bcdot\partial_\mu$.
The scalar-field two-form $\sss$ is defined owing to the standard basis as
\begin{align}
\sss^a&:=d\varphi\wedge\eee^a
=\left(\partial_\bcdot\varphi\right)\eee^\bcdot\wedge\eee^a.
\end{align}
The physical dimensions of the scalar field are set to the reciprocal length in this study.
Consequently, the Lagrangian form is dimensionless: 
\begin{align}
\left[\varphi\right]=L^{-1}\rightarrow \left\{
\begin{array}{ccc}
\left[\iota_a\sss^b\right]&=L^{-1}\\
\left[V\right]&=L^{-4}
\end{array}
\right.
\rightarrow\hspace{.3em}\left[ \LLL_S \right]=1,
\end{align}
where $[\bullet]$ denotes the physical dimensions of object $\bullet$ and $L$ is a physical dimension of length.
On the other hand, the gravitational Lagrangian has dimensions of squared length $\left[\LLL_G\right]=L^2$.
The action integral $\I$ can be defined as
\begin{align}
\I&:=\int(\LLL_G+\hbar\LLL_S).
\end{align}
By requiring a stationary condition on the variation of the action with respect to the vierbein form as $\delta_\eee\I=0$, the Euler-Lagrange equation becomes
\begin{align}
\frac{1}{2}\epsilon_{a\bcdots\bcdot}\RRR^\bcdots\wedge\eee^\bcdot&=-\kappa\hbar~\ttt_a,\label{einstein}
\end{align}
where $\ttt_a$ is the energy-momentum three-form of the scalar field, which can be represented as
\begin{align}
\ttt_a&=-\frac{1}{3!}\epsilon_{a\bcdot\bcdots}(\iota_\star\sss^\bcdot)\wedge(\iota^\star\sss^\bcdot)\wedge\eee^\bcdot+V(\varphi)\VVV_a.
\end{align}
Here, Roman indices are raised and lowered via the Lorentz metric tensor, e.g. $\iota^a\sss^b=\eta^{a\bcdot}\iota_\bcdot\sss^b$.
The torsion-less condition and the Klein--Gordon equation can be obtained from $\delta_{(\www,d\www)}\I=0$ and $\delta_{(\varphi,d\varphi)}\I=0$, respectively.

The Lagrangian (\ref{LagrangianCF}) can be expressed using a trivial frame vector on $\M$ or $\MM$ as
\begin{align}
(\ref{LagrangianCF})
&=\vvv\left(
\frac{1}{2}
\eta^\bcdots\partial_\bcdot\varphi~\partial_\bcdot\varphi
-V(\varphi)\right),\nonumber\\
&=\sqrt{-det[{\bm g}]}\left(
\frac{1}{2}g^{\mu_1\mu_2}\partial_{\mu_1}\varphi~\partial_{\mu_2}\varphi
-V(\varphi)
\right)dx^0\wedge dx^1\wedge dx^2\wedge dx^3.\label{LagrangianCF2}
\end{align}
Here, the relation $\iota_a\eee^b=\delta^b_a$ was used. 
The Einstein equation (\ref{einstein})  can be expressed using the standard basis on $\M$  as
\begin{align*}
&~&R_{ab}-\frac{1}{2}\eta_{ab} R=-2\kappa\hbar~T_{ab},\\
&~&T_{ab}=
\partial_{a}\varphi~\partial_{b}\varphi
-\frac{1}{2}\eta_{ab}\partial_\bcdot\varphi~\partial^\bcdot\varphi
+\eta_{ab}V,
\end{align*}
where $R_{ab}$ and $R$ are the Ricci and scalar curvature, respectively.
The energy-momentum tensor  $T_\bullett$ is defined through relation $\ttt_a=\VVV_{\bcdot}T_{~a}^{\bcdot}$.
\end{appendix}
\bibliographystyle{bmc-mathphys} 
\bibliography{GR-WKB}      
\end{document}